\documentclass[a4paper,12pt]{article}
\usepackage{amsfonts}
\usepackage{latexsym}
\usepackage{amsmath}
\usepackage{amssymb}
\usepackage{amssymb}
\usepackage{slashed}
\usepackage{upgreek}
\usepackage{mathrsfs}
\usepackage{bbm}

\usepackage[toc,page]{appendix}

\usepackage{amsthm}
\usepackage[utf8]{inputenc}
\usepackage{graphicx}

\theoremstyle{remark}

\usepackage[utf8]{inputenc}
\usepackage[english]{babel}

\theoremstyle{definition}

\allowdisplaybreaks

\usepackage{relsize}
\usepackage{graphicx}
\usepackage{comment}

\renewcommand{\thefootnote}{\fnsymbol{footnote}}

\def\appendix#1{\addtocounter{section}{1}\setcounter{equation}{0}
	\renewcommand{\thesection}{\Alph{section}}
	\section*{Appendix \thesection\protect\indent \parbox[t]{11.15cm}{#1}}
	\addcontentsline{toc}{section}{Appendix \thesection\ \ \ #1}}

\newcommand{\pp}{=\kern-0.40em{\vert}}

\def\bbe{{\bf{e}}}

\font\mybb=msbm10 at 11pt

\def\bb#1{\hbox{\mybb#1}}

\def\bR {\bb{R}}

\newcommand{\bea}{\begin{eqnarray}}
	\newcommand{\eea}{\end{eqnarray}}

\usepackage[usenames,dvipsnames,svgnames,table]{xcolor}	
\usepackage[normalem]{ulem}				
\usepackage{microtype}
\usepackage[colorlinks, citecolor=blue, linkcolor=blue, urlcolor=Maroon, filecolor=Maroon, linktocpage=true]{hyperref} 

\usepackage{chngcntr}
\counterwithout{equation}{section}

\begin{document}

	\begin{center}
		\vspace*{-1.0cm}
		\begin{flushright}
		\end{flushright}

		
		\vspace{2.0cm} {\Large \bf W-symmetries, anomalies and heterotic backgrounds with  non-compact holonomy } \\[.2cm]
		
		\vskip 2cm
		 G.\,  Papadopoulos and  J.\, Phillips
		\\
		\vskip .6cm


		\begin{small}
			\textit{Department of Mathematics
				\\
				King's College London
				\\
				Strand
				\\
				London WC2R 2LS, UK}
			\\*[.3cm]
			\texttt{george.papadopoulos@kcl.ac.uk}
			\\
			\texttt{jake.phillips@kcl.ac.uk}
		\end{small}
		\\*[.6cm]

	\end{center}

	\vskip 2.5 cm

	\begin{abstract}
\noindent
We determine the algebra of holonomy symmetries of sigma models propagating on supersymmetric heterotic backgrounds with a non-compact holonomy group. We demonstrate that these close as a W-algebra, which  in turn  is specified  by a Lie algebra structure on the space of covariantly constant forms that generate the holonomy symmetries.    In addition, we identify the chiral anomalies associated with these symmetries. We  argue that these anomalies are consistent and can be cancelled up to two loops in the sigma model perturbation theory.

	\end{abstract}

	

	\newpage
	
	\renewcommand{\thefootnote}{\arabic{footnote}}

\section{Introduction}

 It has been known for sometime \cite{phgpw1, phgpw2}, following earlier related work in \cite{odake, dvn}, that $\hat\nabla$-covariantly constant forms on a spacetime $M$ generate symmetries\footnote{As the existence of $\hat\nabla$-covariantly constant forms on a spacetime implies the reduction of the holonomy of $\hat\nabla$ to a subgroup of $SO(9,1)$, the associated symmetries of  probes are refereed to as holonomy symmetries.} in heterotic probes propagating on $M$, where $\hat\nabla$ is a metric connection with torsion given by the 3-form field strength, $H$, of heterotic  theory, see (\ref{hnabla}). The algebra of these symmetries  is a W-algebra \cite{phgpw2}; for other applications,  see e.g. \cite{vafa, jfof, mg}.   Typically, the heterotic probes  are  (1,0)-supersymmetric sigma models \cite{ewch} that exhibit a metric, a 2-form and  gauge couplings, and  have as a target space the spacetime $M$.    A class of heterotic backgrounds with such $\hat\nabla$-covariantly constant forms are those that preserve a fraction of the spacetime supersymmetry --  the $\hat\nabla$-covariantly constant forms are the (Killing spinor) form bilinears.  The geometry of all supersymmetric heterotic backgrounds has been investigated in \cite{uggp1, uggp2} and the form bilinears have been identified; for a review see \cite{rev}.  There are two  classes of supersymmetric heterotic backgrounds which are distinguished by whether the holonomy group of $\hat\nabla$ is either compact or non-compact. The symmetries of heterotic probes propagating on supersymmetric backgrounds with a compact holonomy group have been investigated in \cite{lggpepb}. In particular, it has been demonstrated that the algebra of symmetries closes as a W-algebra\footnote{The structure constants of the algebra depend on the conserved currents of the symmetries.}, provided that suitable additional symmetries are included. In addition, it has been shown that  the chiral anomalies of these symmetries are consistent and can be cancelled up to two loops in sigma model perturbation theory.

The purpose of this paper is to explore the holonomy symmetries of sigma models on supersymmetric heterotic backgrounds with a non-compact holonomy group. The non-compact holonomy groups that arise are semi-direct products of a compact group, $G$, with $\bR^8$, $G\ltimes\bR^8$,  where $G$ is one of the groups listed in (\ref{hol}).   The distinguished feature of supersymmetric heterotic backgrounds with non-compact holonomy group is that they admit a null $\hat\nabla$-covariantly constant 1-form bilinear $K$
and the remaining form bilinears, $L$, are null along $K$, i.e. they satisfy (\ref{nk}). First, we demonstrate that the commutator of two holonomy symmetries generated by null $\hat\nabla$-covariantly constant forms closes as a W-algebra, see (\ref{walg}).  To describe the commutator in more detail, we find that the space of null forms along $K$ on $M$,  $\Omega^*_K(M)$, admits  a generalisation of the exterior product, $\curlywedge$, and the inner product, $\bar\curlywedge$, operations.  In particular, we find that the commutator of two holonomy symmetries generated by the null forms $L$ and $M$ closes on transformations generated by $K$ and $L\bar\curlywedge M$. Moreover, if $L$ and $M$ are $\hat\nabla$-covariantly constant, then $L\curlywedge M$ and $L\bar\curlywedge M$ are $\hat\nabla$-covariantly constant. Thus the space of all null
$\hat\nabla$-covariantly constant forms, $\Omega^*_{\hat\nabla}(M)$, is closed under these two operations. In fact, $\Omega^*_{\hat\nabla}(M)$ is a Lie algebra with Lie bracket operation  $\bar\curlywedge$, which underpins the W-algebra of holonomy symmetries. As a result, the W-algebra of holonomy symmetries is completely determined from the Lie algebra $\Omega^*_{\hat\nabla}(M)$.

Therefore, to explore the W-algebra of holonomy symmetries it suffices to determine  the Lie algebra structure of $\Omega^*_{\hat\nabla}(M)$. For this is  useful to first identify the Lie algebra of {\it fundamental forms}, $\mathfrak{f}$. For this,  first consider the $\bar\curlywedge$-closure of a (minimal) collection
of null forms on $M$ whose covariant constancy condition determines  the holonomy of the connection $\hat\nabla$.  As $K$ $\bar\curlywedge$-commutes with all other elements of $\Omega^*_{\hat\nabla}(M)$, it is convenient to exclude $K$ as an element of  $\mathfrak{f}$.  Then, $\Omega^*_{\hat\nabla}(M)$ is generated from $\mathfrak{f}$  upon taking $\curlywedge$-products of the elements of $\mathfrak{f}$ and including $K$. The Lie algebra structure of $\Omega^*_{\hat\nabla}(M)$ is unravelled after decomposing $\Omega^*_{\hat\nabla}(M)$ in representations of $\mathfrak{f}$.

After establishing some general results on the Lie algebra structure of $\Omega^*_{\hat\nabla}(M)$, a more detailed investigation is ensued on the Lie algebra structure of $\mathfrak{f}$ and $\Omega^*_{\hat\nabla}(M)$ for each background with holonomy listed in (\ref{hol}). The Lie algebras $\mathfrak{f}$ are tabulated in table 1. Moreover, the Lie algebra structure of $\Omega^*_{\hat\nabla}(M)$ is determined in each case.

The investigation of chiral anomalies in sigma models has a long history \cite{gmpn, laggin, bny, sen, chpt2, phgpa}.  More recently, the chiral anomalies of holonomy symmetries for sigma models with Euclidean signature manifolds as target spaces    have been investigated in \cite{dlossa} and those of sigma models with supersymmetric heterotic backgrounds with compact holonomy group as target spaces have been explored  in \cite{lggpepb}.
 After determining  the W-algebra of holonomy symmetries of sigma models on supersymmetric heterotic backgrounds with non-compact holonomy groups, we find the associated chiral anomalies using Wess-Zumino consistency conditions \cite{zumino}. We demonstrate that the anomalies are consistent up to at least two loops in sigma model perturbation theory.  At the same loop level, we find that the anomalies can be cancelled after an appropriate quantum correction to the covariantly constant forms, which is consistent with the Green-Schwarz anomaly cancellation mechanism \cite{mgjs}.

This paper is organised as follows. In section 2, after defining the operations $\curlywedge$ and $\bar\curlywedge$ on $\Omega^*_K(M)$ and $\Omega^*_{\hat\nabla}(M)$ and summarising some of the geometric properties of heterotic backgrounds with non-compact holonomy group, we determine the commutator (\ref{walg}) of two holonomy symmetries.  Moreover, we argue that the W-algebra of holonomy symmetries is underpinned by the Lie algebras  $\mathfrak{f}$ and $\Omega^*_{\hat\nabla}(M)$.  In section 3, first we establish some general properties of the Lie algebra structure on $\Omega^*_{\hat\nabla}(M)$ and then determine the Lie algebras $\mathfrak{f}$ and $\Omega^*_{\hat\nabla}(M)$ for each of the backgrounds with holonomy group $G\ltimes \bR^8$,  where $G$ listed in equation (\ref{hol}).  In section 4, we give the anomalies of these holonomy symmetries, prove that they are consistent up to at least two-loops and discuss their cancellation, and in section 5 we give our conclusions.

 \section{Geometry and sigma model symmetries}

 \subsection{Geometry of backgrounds with non-compact holonomy}

 As it has already been mentioned, the geometry of a  spacetime $M$, with metric $g$, of supersymmetric heterotic backgrounds can be characterised by the holonomy group\footnote{We assume that the structure group of the spacetime $M$ reduces to a subgroup of the connected component $SO(9,1)$ of the Lorentz group. $M$ is spin manifold and  we always refer to the reduced holonomy group. } of the metric connection, $\hat\nabla$, with skew-symmetric torsion given by the 3-form field strength $H$ of the theory.  In particular, one has that
 \bea
 \hat\nabla_\mu X^\nu=\nabla_\mu X^\nu+{1\over2} H^\nu{}_{\mu\rho} X^\rho~,
 \label{hnabla}
  \eea
  where $\nabla$ is the Levi-Civita connection of  $g$,  $X$ is a vector field on $M$ and  $\mu, \nu, \rho=0, \dots, 9$. For supersymmetric backgrounds, the holonomy group of $\hat\nabla$ is a subgroup of the isotropy group of Killing spinors in $\mathrm{Spin}(9,1)$. This is  either compact or non-compact. The non-compact holonomy groups that can occur  are  $G\ltimes \bR^8$ with $G$ given by one of the following groups \cite{uggp1, uggp2}
 \bea
 &&\mathrm{Spin}(7)~(1)~,~~~SU(4)~(2)~,~~~Sp(2)~(3)~,~~~\times^2Sp(1)~(4)~,~~~
 \cr
 &&Sp(1)~(5)~,~~~U(1)~(6)~,~~~\{1\}~(8)~,
 \label{hol}
 \eea
 where in parenthesis is the number of Killing spinors for each case.  The form bilinears, which are $\hat\nabla$-covariantly constant by construction, include a null 1-form denoted by $K$. The 1-form $K$ is no-where vanishing on the spacetime and is Killing. The remaining form bilinears $L$ are also null along $K$, i.e. they satisfy 
 \bea
 K\wedge L=i_K L=0~,
 \label{nk}
 \eea
 where $i_K$ denotes the inner derivation\footnote{In general, the inner derivation of a $n$-form $N$ with respect to a vector $(m-1)$-form $M$ is $i_M N\equiv M\bar\wedge N\equiv {1\over (m-1)! (n-1)!} M^\nu{}_{\lambda_1\dots \lambda_{m-1}} N_{\nu \lambda_{m}\dots \lambda_{m+n-2}} dx^{\lambda_1\dots \lambda_{m+n-2}}$. We shall use both notations $i_M$ and $M\bar\wedge$ to denote the inner derivation at convenience.} of $L$ with respect to $K$ now viewed as a vector field -- the index is raised with the spacetime metric $g$.

  Let us denote the space of null forms of $M$ along $K$ with   $\Omega^*_K(M)$.
  One way to describe the elements of $\Omega^*_K(M)$, and so the form bilinears,  is to introduce a  local pseudo-orthonormal null co-frame $(\bbe^-, \bbe^+, \bbe^i)$ on the spacetime with $\bbe^-=K$, i.e. $g=2 \bbe^+\bbe^-+\delta_{ij}\bbe^i \bbe^j$.  Then the conditions\footnote{In the dual frame $(\bbe_-, \bbe_+, \bbe_i)$   to the  co-frame $(\bbe^-, \bbe^+, \bbe^i)$, the vector field $K=\bbe^\mu_+\partial_\mu$.}  $K\wedge L=i_KL=0$  can be solved to yield
\bea
L={1\over \ell!}\,L_{-i_1\dots i_\ell}\, \bbe^-\wedge \bbe^{i_1}\wedge \dots \wedge \bbe^{i_\ell}\equiv {1\over \ell!}\,L_{i_1\dots i_\ell}\, \bbe^{-i_1 \dots i_\ell}~,
\label{exL}
\eea
i.e. the only non-vanishing components of $L$ are $L_{-i_1\dots i_\ell}\equiv L_{i_1\dots i_\ell}$, where in the last expression we have simplified the notation for the wedge product of co-frame 1-forms.  Note that to establish (\ref{exL}), we have used that $K$ is no-where vanishing on the spacetime.

Typically for form bilinears, the  components of $L$ can be identified with those of the usual fundamental
forms of $G$-structures in 8 dimensions, up to an equivalence that will be discussed below, where $G$ is given in (\ref{hol}).  For example if $G=\mathrm{Spin}(7)$, the components of $L$ on an open set are identified with those of the fundamental self-dual 4-form of the group. Of course  $L$ depends on all coordinates of $M$.

Prompted by this, we associate to every $L\in \Omega^*_K(M)$ a locally defined form $\tilde L_\alpha$ for each open set, $U_\alpha$, of $M$ such that
\bea
\tilde L_\alpha={1\over \ell!}\, L_{i_1\dots i_\ell}  \bbe^{i_1\dots i_\ell}~,
\eea
i.e. $\tilde L_{i_1\dots i_\ell}= L_{i_1\dots i_\ell}$.  The form $\tilde L=\{\tilde L_\alpha\}$  is not a globally defined  on $M$. Instead at the intersection, $U_\alpha\cap U_\beta$, of two open sets $U_\alpha$ and $U_\beta$,
\bea
\tilde L_\alpha= \tilde L_\beta+ \bbe^-\wedge \tilde N_{\alpha\beta}~,
\eea
where $\tilde N_{\alpha\beta}={1\over (\ell-1)!}N_{i_1\dots i_{\ell-1}}  \bbe^{i_1\dots i_{\ell-1}}$. To establish this, we have used that $L_\alpha=\bbe^- \wedge \tilde L_\alpha=\bbe^- \wedge \tilde L_\beta=L_\beta$ on $U_\alpha\cap U_\beta$ and that $\bbe^-$ is no-where vanishing on the spacetime.  In the following, the above relation between $L$ and $\tilde L$ will be referred to as $ L$ is {\it represented} by $\tilde L$ or equivalently $\tilde L$ {\it represents} $L$.

 We shall use the observations we have made above to define two algebraic operations on $\Omega^*_K(M)$.  Indeed given $L,M\in \Omega^*_K(M)$, we define
\bea
&&{L\bar{\curlywedge} M}\equiv \bbe^-\wedge i_{\tilde L} \tilde M={1\over (\ell-1)! (m-1)!} \,  L^j{}_{i_1\dots i_{\ell-1}}  M_{j i_\ell \dots i_{\ell+m-2}}\,  \bbe^{-i_1\dots i_{\ell+m-2}}~,
\cr
&&L\curlywedge M\equiv L\wedge \tilde M= \bbe^-\wedge \tilde L\wedge \tilde M~,~~~
\label{twoop}
\eea
where we have used $\tilde L$ and $\tilde M$ that represent $L$ and $M$, respectively.
Although both operations\footnote{Note that if $L,M\in \Omega^*_K$, then $L\bar\wedge M=L\wedge M=0$.} $\bar{\curlywedge}$ and $\curlywedge$ are defined using local data, $L\bar{\curlywedge} M$ and $L\curlywedge M$ are globally defined forms on the spacetime. Moreover, if $L$ and $M$ are $\hat\nabla$-covariantly constant, then $L\curlywedge M$ and $L\bar{\curlywedge} M$ are $\hat\nabla$-covariantly constant as well. So clearly, the two operations described in (\ref{twoop}) can be used to construct new  $\hat\nabla$-covariantly constant forms from old ones.

Let us denote with, $\Omega^*_{\hat\nabla}(M)$, the vector space spanned by all null along $K$ $\hat\nabla$-covariantly constant forms of a spacetime $M$. As all such forms have odd degree for the backgrounds we shall be investigating, $\Omega^*_{\hat\nabla}(M)$ with bracket operation  $\bar{\curlywedge}$ is a Lie algebra. As we shall demonstrate, the algebra of holonomy symmetries of sigma models with target spaces given by  supersymmetric heterotic backgrounds with non-compact holonomy groups is determined by the Lie algebra $\Omega^*_{\hat\nabla}(M)$. Therefore, the Lie algebra structure of $\Omega^*_{\hat\nabla}(M)$ is of interest.  Some of it is easily unravelled. First, $\Omega^1_{\hat\nabla}(M)$ is spanned by the null 1-form $K$ which commutes with all the remaining elements of $\Omega^*_{\hat\nabla}(M)$. In addition, $\Omega^9_{\hat\nabla}(M)$ is spanned by the Hodge dual form $E$ of $K$, $E=\star K$.  Again $E$ commutes with the rest of the elements of $\Omega^*_{\hat\nabla}(M)$.  Therefore both $K$ and $E$ are in the centre of $\Omega^*_{\hat\nabla}(M)$.

To further explore  the Lie algebra structure of $\Omega^*_{\hat\nabla}(M)$, one defines the Lie algebra of fundamental forms $\mathfrak{f}$ as described in the introduction. This is a Lie subalgebra, $\mathfrak{f}$, of $\Omega^*_{\hat\nabla}(M)$.  In the supersymmetric backgrounds considered here, $\mathfrak{f}$ is spanned by form bilinears\footnote{An exception are the  backgrounds with holonomy $U(1)\ltimes\bR^8$; this will be explained later.}.
 In all examples that we shall investigate,  $\Omega^*_{\hat\nabla}(M)$ is generated by $\mathfrak{f}$ upon taking the $\curlywedge$ product of elements of $\mathfrak{f}$ and including $K$.  Moreover, it turns out that, apart from backgrounds with holonomy $\mathrm{Spin}(7)\ltimes \bR^8$ and $SU(4)\ltimes \bR^8$, $\mathfrak{f}$ is generated by 3-forms.  In all such cases, the algebraic structure of $\Omega^*_{\hat\nabla}(M)$, and so that of the symmetries of the sigma model, can be unravelled by decomposing $\Omega^*_{\hat\nabla}(M)$ into irreducible representations of $\mathfrak{f}$.  The complete structure is presented later on a case by case basis.

 Another ingredient needed in the analysis  is a description of the geometry of supersymmetric heterotic backgrounds with non-compact holonomy \cite{uggp1, uggp2}. For our purposes, a brief outline suffices. In particular,
 after solving the Killing spinor equations of heterotic theory, the fields  can be expressed as
 \bea
g&=&2\bbe^+ \bbe^-+\delta_{ij}\bbe^i\bbe^j~,
\cr
H&=&H_{+-i} \bbe^{+-i}+{1\over 2} H_{+ij} \bbe^{+ij}+{1\over2} H_{-ij} \bbe^{-ij}+{1\over3!} H_{ijk} \bbe^{ijk}
\cr 
&=&d\bbe^-\wedge \bbe^++{1\over2} H_{-ij} \bbe^{-ij}+{1\over3!} H_{ijk} \bbe^{ijk}~,
\cr
F&=&F_{-i} \bbe^{-i}+{1\over2} F_{ij} \bbe^{ij}~,
\label{mh}
\eea
where $F$ is the curvature of the gauge sector and we have suppressed the gauge indices.
Most of the components of $H$ given above are determined in terms the metric and the form bilinears of the theory \cite{uggp1, uggp2}.  Though, there is no need to give a detailed description. However, it is significant that the Killing spinor equations imply that $i_KH$ satisfies
\bea
i_L(i_K H)=0 \Longleftrightarrow H_{+\nu [\mu_1} L^\nu{}_{\mu_2\dots \mu_{\ell+1}]}=0~,
\label{invcon1}
\eea
for all fundamental forms $L$ of the holonomy group $G\ltimes\bR^8$.
Using this and the $\hat\nabla$-covariantly constancy of $L$, one can show that the Lie derivative of $L$ with respect to $K$ vanishes
\bea
{\cal L}_K L=0~.
\label{liedev}
\eea
 Therefore, all fundamental forms are invariant under the action of the vector field $K$. Furthermore, the gaugino Killing spinor equation implies that $i_K F=i_L F=0$. Also (\ref{liedev})  holds  for all $L\in \Omega^*_{\hat\nabla}(M)$, i.e. not only for the fundamental forms,  provided that $i_L(i_K H)=0$.

\subsection{Holonomy symmetries of chiral sigma models}

A description of chiral 2-dimensional sigma models suitable for the analysis that follows has already been presented in \cite{lggpepb}. So we shall be brief. The  fields of the sigma model are  maps, $X$, from the worldsheet superspace $\Xi^{2|1}$, with coordinates $(\sigma^=, \sigma^{\pp}, \theta^+)$, into a spacetime $M$ and Grassmannian odd sections, $\psi$, of a vector bundle $S_-\otimes X^*E$ over $\Xi^{2|1}$.  Here $S_-$ is the anti-chiral spinor bundle over $\Xi^{2|1}$ and $E$ is a vector bundle over $M$. An action   for these fields \cite{ewch} is
\bea
S=-i \int d^2\sigma d\theta^+  \Big ((g_{\mu\nu}+b_{\mu\nu}) D_+X^\mu \partial_=X^\nu+i h_{\mathrm{a}\mathrm{b}} \psi_-^\mathrm{a}{\cal D}_+\psi_-^\mathrm{b}\Big)~,
\label{act1}
\eea
where $g$ is a spacetime metric, $b$ is a locally defined 2-form on $M$, such that $H=db$ is a globally defined 3-form. In addition,  $D_+^2=i\partial_{\pp}$ and $h$ is a fibre metric on $E$,
\bea
{\cal D}_+\psi_-^{\mathrm{a}}=D_+\psi_-^{\mathrm{a}}+ D_+X^\mu \Omega_\mu{}^{\mathrm{a}}{}_{\mathrm{b}} \psi_-^{\mathrm{b}}~,
\eea
where $\Omega$ is a connection on $E$  and ${\cal D}_\mu h_{\mathrm{a}\mathrm{b}}=0$. We shall refer to the part of the action with couplings $h$ and ${\cal D}$ as the gauge sector of the theory. Note that
\bea
\delta S= -i \int d^2\sigma d\theta^+ \big(\delta X^\mu {\cal S}_\mu+\Delta \psi_-^{\mathrm{a}} {\cal S}_{\mathrm{a}}\big)~,
\eea
where  
\bea
{\cal S}_\mu=-2 g_{\mu\nu} \hat\nabla_= D_+ X^\nu-i  \psi_-^{\mathrm{a}} \psi_-^{\mathrm{b}} D_+X^\nu F_{\mu\nu \mathrm{a}\mathrm{b}}~,~~
{\cal S}^{\mathrm{a}}=2i{\cal D}_+\psi_-^{\mathrm{a}}~,
\label{feqns}
\eea
are the field equations, $
\Delta \psi_-^{\mathrm{a}} \equiv \delta \psi_-^{\mathrm{a}} + \delta X^\mu \Omega_\mu{}^{\mathrm{a}}{}_{\mathrm{b}} \psi_-^{\mathrm{b}}$ and   $F$ is the curvature of the connection $\Omega$ of the gauge sector.

Before we proceed to describe the holonomy symmetries we shall mention two sigma model symmetries that are relevant in the analysis of anomalies.
First, in  the background field method of quantising the theory \cite{honer, mukhi, friedan, phgpks, blasi}, it is convenient to express the quantum field in a frame basis. In such a case, if we write the metric as $g_{\mu\nu}= \eta_{AB} \bbe^A_\mu \bbe^B_\nu$, then the action of infinitesimal spacetime frame rotations will be
\bea
\delta_\ell \bbe^A_\mu= \ell^A{}_B  \bbe^B_\mu~,~~~~\delta_\ell \omega_\mu{}^A{}_B=-\partial_\mu \ell^A{}_B+ \ell^A{}_C\,  \omega_\mu{}^C{}_B -\omega_\mu{}^A{}_C\,  \ell^C{}_B~,
\label{ftran2}
\eea
where $\ell$ is the infinitesimal parameter and  $\omega$ is a frame connection of the tangent bundle which we shall always assume that preserves the spacetime metric.

Moreover, the fields and coupling constants of the gauge sector transform under  infinitesimal gauge transformations  as
\bea
&&\delta_u \psi^{\mathrm{a}}_-= u^{\mathrm{a}}{}_{\mathrm{b}}\psi^{\mathrm{b}}_-~,~~~\delta_u\Omega_\mu{}^{\mathrm{a}}{}_{\mathrm{b}}=-\partial_\mu u^{\mathrm{a}}{}_{\mathrm{b}}+ u^{\mathrm{a}}{}_{\mathrm{c}}\,  \Omega_\mu{}^{\mathrm{c}}{}_{\mathrm{b}} -\Omega_\mu{}^{\mathrm{a}}{}_{\mathrm{c}}\,  u^{\mathrm{c}}{}_{\mathrm{b}}~,
\cr
&&\delta_u h_{\mathrm{a}\mathrm{b}}=-  u^{\mathrm{c}}{}_{\mathrm{a}} h_{\mathrm{c}\mathrm{b}}- h_{\mathrm{a}\mathrm{c}}    u^{\mathrm{c}}{}_{\mathrm{b}}~,
\label{ftran1}
\eea
where $u$ is the infinitesimal parameter and  the remaining fields and couplings of the theory remain inert.

The commutator of two spacetime frame rotations  (\ref{ftran2}) is $[\delta_\ell, \delta_{\ell'}]=\delta_{[\ell, \ell']}$, where $[\cdot, \cdot]$ is the usual commutator of two matrices.
 A similar result holds for the commutator of two  gauge transformations (\ref{ftran1}). In addition to these, there is a gauge symmetry
$\delta b_{\mu\nu}=(dm)_{\mu\nu}$
associated with the 2-form gauge potential $b$, where $m$ is a 1-form on the spacetime.

To describe the holonomy symmetries of the sigma model action (\ref{act1}), let $L$  be a  $(\ell+1)$-form  on the sigma model target space $M$ and consider the infinitesimal transformation
\bea
\delta_L X^\mu=a_L L^\mu{}_{\lambda_1\dots \lambda_\ell} D_+X^{\lambda_1}\dots D_+X^{\lambda_\ell}\equiv a_L L^\mu{}_L D_+X^L~,~~~\Delta_L\psi_-^\mathrm{a}=0~,
\label{Gsym}
\eea
where $a_L$ is the parameter of the transformation, chosen such that $\delta_L X^\mu$ is even under Grassmannian parity. The index $L$ is the multi-index $L=\lambda_1 \dots \lambda_\ell$ and $D_+X^L=D_+X^{\lambda_1}\cdots D_+X^{\lambda_\ell}$.  Such a transformation \cite{phgpw1, phgpw2} leaves the action  (\ref{act1}) invariant provided that
\bea
\hat\nabla_{\nu} L_{\lambda_1\dots\lambda_{\ell+1}}=0~,~~~F_{\nu[\lambda_1} L^\nu{}_{\lambda_2\dots \lambda_{\ell+1}]}=0~,
\label{invcon}
\eea
i.e. $L$ is $\hat\nabla$-covariantly constant and $i_LF=0$. For form bilinears the former condition is  satisfied as a consequence of the gravitino Killing spinor equations while  the latter condition is a consequence of the gaugino Killing spinor equation.
 Moreover, the parameter $a_L$ satisfies $\partial_=a_L=0$, i.e. that $a_L=a_L(x^{\pp}, \theta^+)$.

The commutator of two transformations (\ref{Gsym}) on the field $X$ has been explored in detail in  \cite{sven, gpph}. Here, we shall summarise some of the key formulae. The commutator of two  transformations (\ref{Gsym}) on the field\footnote{In all cases considered here,  the commutator of two holonomy symmetries on the field $\psi$ gives rise to a transformation with $\Delta_{LM}\psi=0$. So it will not be further investigated. } $X$ generated by the $(\ell+1)$-form $L$ and the $(m+1)$-form $M$  can be written as
\bea
[\delta_L, \delta_M]X^\mu= \delta_{LM}^{(1)} X^\mu+\delta_{LM}^{(2)} X^\mu+\delta_{LM}^{(3)} X^\mu~,
\label{comm}
\eea
with
\bea
\delta_{LM}^{(1)} X^\mu=a_M a_L N(L,M)^\mu{}_{LM} D_+X^{LM}~,
\eea
\bea
\delta_{LM}^{(2)} X_\mu&=&\big (-m a_M D_+a_L (L\cdot M)_{\nu L_2, \mu M_2}
\cr
&&
\qquad\qquad
+ \ell (-1)^{(\ell+1) (m+1)} a_L D_+ a_M (L\cdot M)_{\mu L_2,\nu M_2}\big)
 D_+X^{\nu L_2M_2}~,
\eea
and
\bea
\delta_{LM}^{(3)} X_\mu=-2i \ell m (-1)^\ell a_M a_L (L\cdot M)_{(\mu|L_2|, \nu )M_2} \partial_{\pp}X^\nu D_+X^{L_2M_2}~,
\eea
where
\bea
(L\cdot M)_{\lambda L_2,\mu M_2}=L_{\rho \lambda[L_2} M^\rho{}_{|\mu| M_2]}~.
\eea
The multi-index $M$ stands for $M=\mu_1\dots\mu_m$ while the multi-indices $L_2$ and $M_2$ stand for $L_2=\lambda_2\dots\lambda_\ell$ and $M_2=\mu_2\dots \mu_m$, respectively,
and $N(L,M)$ is the Nijenhuis tensor of $L$ and $M$ which we shall not state here -- it is a generalisation of the standard Nijenhuis tensor of an almost complex structure.  Using that $L$ and $M$ are $\hat\nabla$-covariantly constant, the Nijenhuis tensor can be re-expressed as
\bea
N_{\mu LM} dx^{LM}=  \Big(- (\ell+m+1)H_{[\mu|\nu\rho|} L^\nu{}_L M^\rho{}_{M]}+\ell m H^{\rho}{}_{\lambda_1\mu_1} ( L\cdot M)_{(\mu|L_2|, \rho)M_2}\Big)\, dx^{LM} ~,
\eea
in terms of $H$.
 The conserved current of a symmetry generated by the $(\ell+1)$-form $L$ is
\bea
J_L=L_{\mu_1\dots\mu_{\ell+1}} D_+X^{\mu_1\dots\mu_{\ell+1}}~.
\label{curL}
\eea
It can  easily be seen that $\partial_=J_L=0$ subject to the field equations (\ref{feqns}).


\subsection{The commutator of null holonomy symmetries}

The commutator of two symmetries generated by two $\hat\nabla$-covariantly constant forms $L$ and $M$ is significantly simplified whenever $L,M\in \Omega^*_{\hat\nabla}(M)$.   But before we state this, one can prove, using   ${\cal L}_K L=0$ and $i_KL=0$, that
\bea
[\delta_K, \delta_L] X^\mu=0~,
\eea
for any  $\hat\nabla$-covariantly constant null along $K$ form $L$.  Therefore, the symmetries generated by the Killing vector $K$ commutes with all other holonomy symmetries.

After some computation, using that $L,M \in \Omega^*_{\hat\nabla}(M)$,  ${\cal L}_K L={\cal L}_K M=0$ and  the condition (\ref{invcon1}),  the commutator of two symmetries generated by  $L$ and $M$ can be expressed as
\bea
[\delta_L, \delta_M] =\delta_K+ \delta_{{L\bar{\curlywedge} M}}~,
\label{walg}
\eea
with $a_K=-{\ell! m!\over (\ell+m-1)!} D_+(a_L a_M J_{L\bar{\curlywedge} M})$ and $a_{{L\bar{\curlywedge} M}}=- {\ell! m!\over (\ell+m-2)!} a_L a_M\, D_+J_K$,   where ${L\bar{\curlywedge} M}$ is given in (\ref{twoop}) and is a globally defined form on the spacetime. Note that in all cases that we shall be considering, the forms that generate the holonomy symmetries have odd degree, and so the parameters $a_L, a_M$ of the transformations have even Grassmannian parity.  As the structure constants of the algebra depend on the conserved currents of the associated symmetries, the algebra of variations closes as a W-algebra.

Furthermore, the W-algebra (\ref{walg})  closes to transformations generated by $K$ and $L\bar{\curlywedge} M$. As $L\bar{\curlywedge} M\in \Omega^*_{\hat\nabla}(M)$ for all $L,M \in \Omega^*_{\hat\nabla}(M)$  and $i_{L\bar{\curlywedge} M} F=0$, $L\bar{\curlywedge} M$ generates a new symmetry for the sigma model action (\ref{act1}). Therefore, the commutator (\ref{walg}) of the W-algebra of holonomy symmetries is determined by the Lie algebra $\Omega^*_{\hat\nabla}(M)$ with bracket operation $\bar{\curlywedge}$. So to determine the W-algebra of holonomy symmetries, it remains to identify the Lie algebra $\Omega^*_{\hat\nabla}(M)$ for each of the holonomy groups (\ref{hol}).


\section{The W-algebra of null holonomy symmetries}

\subsection{The W-algebra of null Spin(7) and SU(4) symmetries}

In both these cases, the space of fundamental forms $\mathfrak{f}$ contains forms of degree greater than three. This distinguishes them from the other holonomy groups stated in (\ref{hol}) and so they are separately investigated.

\subsubsection{Spin(7)}

The holonomy symmetries in this case are generated by $K$ and a null 5-form spinor bilinear $L$ which is represented by the usual self-dual, $\mathrm{Spin}(7)$ invariant, fundamental 4-form $\tilde L$. Using the algebraic properties of $\tilde L$, one can demonstrate that $L\bar{\curlywedge} L=0$. Therefore $\mathfrak{f}=\bR\langle L\rangle$.

 Moreover $L{\curlywedge} L$ is proportional to $E=\star K$.  As $K, E$ are in the centre of $\Omega^*_{\hat\nabla}(M)$, one concludes that $\Omega^*_{\hat\nabla}(M)=\bR^3\langle K, L, E\rangle$ is abelian. As a result, the W-algebra of symmetries (\ref{walg}) is abelian as well.

\subsubsection{SU(4)}

To describe  the geometry,  introduce a pseudo-hermitian co-frame $(\bbe^-, \bbe^+, \bbe^\alpha, \bbe^{\bar\alpha})$, $\alpha=1,\dots,4$, on the spacetime. In this co-frame, the metric is expressed as $g=2 \bbe^-\bbe^++2\delta_{\alpha\bar\beta} \bbe^\alpha \bbe^{\bar\beta}$ and the generators  of the fundamental  forms, other than $K=\bbe^-$,  are given by
\bea
&&I={1\over2} I_{ij} \bbe^-\wedge \bbe^{ij}\equiv -i \delta_{\alpha\bar\beta} \bbe^{-\alpha\bar\beta}~,~~~
\cr
&&
L_1={1\over4!} (L_1)_{i_1\dots i_4}  \bbe^{-i_1\dots i_4}\equiv {1\over4!} \epsilon_{\alpha_1\dots \alpha_4}  \bbe^{-\alpha_1\dots \alpha_4}+{1\over4!} \epsilon_{\bar\alpha_1\dots \bar\alpha_4}  \bbe^{-\bar\alpha_1\dots \bar\alpha_4}~,
\cr
&&L_2={1\over4!} (L_2)_{i_1\dots i_4}  \bbe^{-i_1\dots i_4}\equiv -{i\over4!} \epsilon_{\alpha_1\dots \alpha_4}  \bbe^{-\alpha_1\dots \alpha_4}+{i\over4!} \epsilon_{\bar\alpha_1\dots \bar\alpha_4}  \bbe^{-\bar\alpha_1\dots \bar\alpha_4}~.
\eea
Note that the conditions on $i_K H$ arising from $i_I i_KH= i_{L_1} i_K H=i_{L_2} i_K H=0$ in the pseudo-hermitian co-frame can be written as
$H_{+\alpha}{}^\alpha=H_{+\alpha\beta}=0$.

A straightforward calculation reveals that
\bea
I\bar{\curlywedge} L_1=-4L_2~,~~~I\bar{\curlywedge} L_2=4L_1~,~~~L_1\bar{\curlywedge} L_2=-{3\over2} {1\over 6^2} I_{i_1i_2} I_{i_3i_4} I_{i_5 i_6}  \bbe^{-i_1\dots i_6}=-{1\over3} \curlywedge^3I~,
\eea
where all the remaining $\bar{\curlywedge}$ operations amongst these form  vanish. Therefore, closure of the Lie algebra of the fundamental forms $\mathfrak{f}$
requires the introduction of a new generator $\curlywedge^3I$ that commutes with $I$, $L_1$ and $L_2$. Therefore, $\curlywedge^3I$ is a central generator. In fact,  $\mathfrak{f}=\hat {\mathfrak{e}}(2)$,  where $\hat {\mathfrak{e}}(2)$ is the central extension  of the Euclidean algebra, $\mathfrak{e}(2)=\mathfrak{so}(2)\oplus_s\bR^2$, with $I$ the generator of $\mathfrak{so}(2)$ and $L_1, L_2$ the generators of $\bR^2$.

To give an example where the W-algebra of holonomy symmetries is determined from the Lie algebra $\mathfrak{f}$, consider this case with $\mathfrak{f}=\hat {\mathfrak{e}}(2)$.  Then, the  commutators (\ref{walg}) read
\bea
&&[\delta_I, \delta_{L_1}]= {8\over5} D_+(a_L a_M J_{L_2})\delta_K+ 8  a_L a_M\, D_+J_K \,\delta_{{L_2}}
\cr
&&[\delta_I, \delta_{L_2}]= -{8\over5} D_+(a_L a_M J_{L_1})\delta_K- 8  a_L a_M\, D_+J_K \,\delta_{{L_1}}
\cr
&&
[\delta_{L_1}, \delta_{L_2}]={4\over15}\Big({1\over7} D_+(a_L a_M J_{\curlywedge^3I})\delta_K+   a_L a_M\, D_+J_K \,\delta_{{\curlywedge^3I}}\big)~,
\eea
with the remaining commutators to vanish. Notice that  the structure constants depend on the currents $J_K$, $J_{L_1}$,  $J_{L_2}$ and $J_{\curlywedge^3I}$.

In addition to the generators of $\mathfrak{f}$ and  $K$, $\Omega^*_{\hat\nabla}(M)$ contains two more generators given by $\curlywedge^2 I$ and $\curlywedge^4 I$. The latter is proportional to $E$. In fact $L_1$, $L_2$ and $\curlywedge^2 I$ span the space of 5-form bilinears. Moreover,  $L_1\curlywedge L_1$ and $L_2\curlywedge L_2$ are proportional to $E$ and $L_1\curlywedge L_2=0$. It turns out that $\curlywedge^2 I$ and $\curlywedge^4 I$ commute  amongst themselves as well as with the generators of $\hat {\mathfrak{e}}(2)$. As a result, one concludes  that $\Omega^*_{\hat\nabla}(M)=\hat {\mathfrak{e}}(2)\oplus \bR^3\langle K, \curlywedge^2 I, E\rangle$.

\subsection{The W-algebra of remaining null  holonomy symmetries}\label{tpt}

 For all the remaining backgrounds that are investigated below, $\mathfrak{f}$ is spanned by 3-forms. In addition,  $\Omega^3_{\hat\nabla}(M)=\mathfrak{f}$ and $\Omega^7_{\hat\nabla}(M)$ is the Hodge dual space of $\mathfrak{f}$, $\Omega^7_{\hat\nabla}(M)=\star \mathfrak{f}$.

To determine the Lie algebra structure of $\Omega^*_{\hat\nabla}(M)$, note that $\Omega^*_{\hat\nabla}(M)$ decomposes into (irreducible) representations of $\mathfrak{f}$. Clearly $\mathfrak{f}$ acts on $\Omega^3_{\hat\nabla}(M)$ with the adjoint representation. It also acts with the same representation on
$\Omega^7_{\hat\nabla}(M)$ as
\bea
L\bar\curlywedge \star M=\star (L \bar\curlywedge M)~,
\eea
for any $L,M\in \Omega^3_{\hat\nabla}(M)$.  A calculation also reveals that
\bea
L\bar\curlywedge M=0~,
\eea
for any $L\in \Omega^5_{\hat\nabla}(M)$ and $M\in \Omega^7_{\hat\nabla}(M)$.  So to determine the Lie algebra structure of $\Omega^*_{\hat\nabla}(M)$,  it remains to compute $\Omega^3_{\hat\nabla}(M)\bar\curlywedge\Omega^5_{\hat\nabla}(M)$ and $\Omega^5_{\hat\nabla}(M)\bar\curlywedge\Omega^5_{\hat\nabla}(M)$. The former will be identified on a case by case basis. In computing the latter, we shall use the  formula
\bea
(L\curlywedge M)\bar\curlywedge (N\curlywedge P)&=&(L\bar\curlywedge N)\curlywedge M \curlywedge P+(L\bar\curlywedge P)\curlywedge M \curlywedge N
\cr
&&+(M\bar\curlywedge N)\curlywedge L \curlywedge P+(M\bar\curlywedge P)\curlywedge L \curlywedge N~,
\label{LMNP}
\eea
 where $L,M,N,P\in \Omega^3_{\hat\nabla}(M)$. For the backgrounds examined below, all the elements of $\Omega^5_{\hat\nabla}(M)$ can be written as the $\curlywedge$-product of two elements in
$\Omega^3_{\hat\nabla}(M)$.

\subsubsection{Sp(2)}

The Lie algebra of fundamental forms $\mathfrak{f}$ is spanned by
\bea
I_1={1\over2} (I_1)_{ij} \bbe^{-ij}~,~~~I_2={1\over2} (I_2)_{ij} \bbe^{-ij}~,~~~I_3={1\over2} (I_3)_{ij} \bbe^{-ij}~,
\label{sp2bi}
\eea
where their components are represented by the usual hypercomplex structure that characterises $Sp(2)$ in eight dimensions.  In particular, $\tilde I_1^2=\tilde I_2^2=-{\bf 1}_{8\times 8}$, $\tilde I_1\tilde I_2+\tilde I_2 \tilde I_1=0$ and $\tilde I_3=\tilde I_1 \tilde I_2$.  It is clear from this that
\bea
I_r\bar\curlywedge I_s=-2 \epsilon_{rs}{}^t I_t~,
\eea
 and so $\mathfrak{f}=\mathfrak{sp}(1)$. Notice that  $\mathfrak{sp}(2)\oplus \mathfrak{sp}(1)$ is a maximal subalgebra of $\mathfrak{so}(8)$.


To describe the Lie algebra structure of $\Omega^*_{\hat\nabla}(M)$,  define
\bea
I_{r_1\dots r_p}\equiv I_{r_1} \curlywedge\dots \curlywedge I_{r_p}~,~~~p=1,2,3,4~.~~~
\eea
Next $\Omega^5_{\hat\nabla}(M)=\bR\langle I_{rs}\rangle$. As $I_{rs}=I_{sr}$ and
\bea
I_r\bar{\curlywedge} I_{s_1\dots s_q}&=&-2q \epsilon_{r(s_1}{}^{t} I_{|t|\dots s_q)}=2q \epsilon_{(s_1|r}{}^{t} I_{t|\dots s_q)}~,
\eea
$\mathfrak{f}=\mathfrak{sp}(1)=\mathfrak{so}(3)$ acts on $\Omega^5_{\hat\nabla}(M)$ with the symmetric product of two vector representations of $\mathfrak{so}(3)$. This decomposes as $\Omega^5_{\hat\nabla}(M)= \Omega^5_{\bf 5}\oplus \Omega^5_{\bf 1}$, where $\Omega^5_{\bf 5}$ is the irreducible symmetric traceless representation and $\Omega^5_{\bf 1}$ is the trivial representation, which is spanned by the Casimir element  $C=I_{11}+I_{22}+I_{33}$. In fact $C$ commutes with all elements of $\Omega^*_{\hat\nabla}(M)$ and is an element of the centre, together with $K$ and $E$.

It remains to compute $\Omega^5_{\hat\nabla}(M)\bar\curlywedge\Omega^5_{\hat\nabla}(M)$. For this, one can use the formula (\ref{LMNP}) to deduce
\bea
I_{r_1r_2}\bar\curlywedge I_{s_1s_2}&=&-2 \epsilon_{r_1s_1}{}^t I_{tr_2s_2}-2 \epsilon_{r_1s_2}{}^t I_{tr_2s_1}-2 \epsilon_{r_2s_1}{}^t I_{tr_1s_2}-2 \epsilon_{r_2s_2}{}^t I_{tr_1s_1}
\cr
&=&-16 (\epsilon_{r_1s_1}{}^t \delta_{r_2s_2}+\epsilon_{r_1s_2}{}^t \delta_{r_2s_1}+\epsilon_{r_2s_1}{}^t \delta_{r_1s_2}+\epsilon_{r_2s_2}{}^t \delta_{r_1s_1}) \star I_t~,
\eea
where we have used
 \bea
 \star I_r={1\over 4!} I_{rrr}~,~~~ \star I_r={1\over 8} I_{ssr}~,~~~r\not=s~,~~~I_{123}=0~,
 \eea
 which lead to
 \bea
 I_{rst}={1\over3}(\delta_{rs} I_{ttt}+\delta_{rt} I_{sss}+\delta_{st} I_{rrr})~.
 \eea
 This together with the results in the beginning of section \ref{tpt} completely determine the Lie algebra structure of $\Omega^*_{\hat\nabla}(M)$ and so the commutators (\ref{walg}) of the associated holonomy symmetries.

\subsubsection{$\times^2$Sp(1)}

In an adapted pseudo-Hermitian co-frame $(\bbe^-, \bbe^+, \bbe^\alpha, \bbe^{\bar\alpha}; \alpha=1,\dots,4)$,  to the $\times^2Sp(1)\ltimes \bR^8$ holonomy, the Lie algebra of fundamental forms  $\mathfrak{f}$ is spanned by
\bea
I_r= {1\over2} (\tilde I_r)_{ij}\, \bbe^{-ij}~,~~~J_r={1\over2}  (\tilde J_r)_{ij}\, \bbe^{-ij}~,
\label{IrJs}
\eea
which are represented by
\bea
&&\tilde I_1= -i(\bbe^{1\bar1}+\bbe^{2\bar2})~,~~~\tilde I_2= \bbe^{12}+\bbe^{\bar 1\bar2}~,~~~\tilde I_3= -i(\bbe^{12}-\bbe^{\bar 1\bar2})~,
\cr
&&\tilde J_1= i(\bbe^{3\bar3}+\bbe^{4\bar4})~,~~~\tilde J_2= \bbe^{34}+\bbe^{\bar 3\bar4}~, \tilde J_3= i(\bbe^{34}-\bbe^{\bar 3\bar4})~,
\eea
 and so $\tilde I_r \tilde I_s=-\delta_{rs} {\bf 1}_{4\times 4}+\epsilon_{rs}{}^t \tilde I_t$ and $\tilde J_r \tilde J_s=-\delta_{rs} {\bf 1}_{4\times 4}+\epsilon_{rs}{}^t \tilde J_t$.   Thus, one has that
\bea
I_r\bar\curlywedge I_s=-2 \epsilon_{rs}{}^t I_t~,~~~J_r\bar\curlywedge J_s=-2 \epsilon_{rs}{}^t J_t~,~~~I_r\bar\curlywedge J_s=0~,
\label{sp1sp1}
\eea
and so  $\mathfrak{f}=\Omega^3_{\hat\nabla}(M)=\oplus^2 \mathfrak{sp}(1)$.  To see this, observe that the action of $\times^2Sp(1)$ on $\bR^8$ can be seen as two copies of the action of
$Sp(1)$ on $\bR^4$ with each copy associated with a hyper-complex structure\footnote{It is conventional that if the (holonomy) group $Sp(1)$ is associated with the hyper-complex structure $(I_1,I_2, I_3)$ on $\bR^4$, then $\mathfrak{sp}(1)$ spans the subspace of (1,1)-forms with respect to $I_1$ on $\bR^4$ which are in addition $I_1$-traceless, i.e. $\mathfrak{sp}(1)$ spans the anti-self dual 2-form while the hyper-complex structure the self-dual 2-forms on $\bR^4$.} on $\bR^4$.

Next define the forms
\bea
&&L\equiv\curlywedge^2I_1=\curlywedge^2I_2=\curlywedge^2I_3 ~,~~~M\equiv\curlywedge^2J_1=\curlywedge^2J_2=\curlywedge^2J_3~,~~~
\cr
&&
N_{rs}\equiv I_r\curlywedge J_s~,~~~R_{r}\equiv I_r\curlywedge M=4\star I_r~,~~~S_{r}\equiv L\curlywedge J_r=4 \star J_r~,~~~
\cr
&&
E={1\over4}L \curlywedge M~.
\label{LMN}
\eea
It can be shown that $\Omega^5_{\hat\nabla}(M)=\bR\langle L, M, N_{rs}\rangle$.  It turns out that $L,M$ together with $K$ and $E$ are in the centre of $\Omega^*_{\hat\nabla}(M)$ while
\bea
I_r\bar{\curlywedge} N_{r's'}=-2\epsilon_{rr'}{}^{t'} N_{t's'}~,~~~J_r\bar{\curlywedge} N_{r's'}=-2\epsilon_{rs'}{}^{t'} N_{r't'}~.
\label{comIN}
\eea
Thus $\Omega^5_{\hat\nabla}(M)$ decomposes under the action of $\mathfrak{f}$ as $\Omega^5_{\hat\nabla}(M)=\Omega^5_{\bf 9}\oplus \bR^2\langle L,M\rangle$, where
$\Omega^5_{\bf 9}$ spanned by $N_{rs}$ is identified with the traceless symmetric product of two vector representations of $\mathfrak{so}(4)=\oplus^2 \mathfrak{sp}(1)$.

To specify the Lie algebra structure of $\Omega^*_{\hat\nabla}(M)$ it remains to determine $\Omega^5_{\hat\nabla}(M)\bar{\curlywedge}\Omega^5_{\hat\nabla}(M)$.
As $L,M$ are in the centre of $\Omega^*_{\hat\nabla}(M)$, a straightforward computation using (\ref{LMNP}) reveals that
\bea
N_{rs}\bar{\curlywedge} N_{r's'}=-2\epsilon_{rr'}{}^{t'} \delta_{ss'} R_{t'}-2\epsilon_{ss'}{}^{t'} \delta_{rr'} S_{t'}~.
\eea
This together with the results in the beginning of section \ref{tpt} completely determine the Lie algebra structure $\Omega^*_{\hat\nabla}(M)$ and so the commutators (\ref{walg}) of the W-algebra of holonomy symmetries.

\subsubsection{Sp(1)}

 In addition to $I_r$ and $J_r$ (\ref{IrJs}) fundamental forms of holonomy $\times^2Sp(1)\ltimes\bR^8$ backgrounds,  the Lie algebra $\mathfrak{f}$ of holonomy $Sp(1)\ltimes\bR^8$ backgrounds contains the  element $A$ which  in the pseudo-hermitian basis of the previous section reads
\bea
A=\bbe^{-13}+\bbe^{-24}+\bbe^{-\bar 1 \bar 3}+\bbe^{-\bar 2 \bar 4}~.
\eea
This is one of the additional  3-form bilinears associated with the holonomy $SU(2)\ltimes\bR^8$ backgrounds.  Next define
\bea
W_r\equiv I_r \bar{\curlywedge} A=-J_r \bar{\curlywedge} A ~.
\eea
Then it turns out that $\mathfrak{f}=\bR\langle I_r, J_s, A, W_t\rangle$.  The non-vanishing commutators are  those already described for backgrounds with holonomy $\times^2 Sp(1)\ltimes\bR^8$ in
(\ref{sp1sp1}) and
\bea
&&I_r\bar{\curlywedge}A=-J_r\bar{\curlywedge} A=W_r~,~~~I_r\bar{\curlywedge} W_s= -\delta_{rs} A-\epsilon_{rs}{}^t W_t~,~~~
\cr
&&J_r\bar{\curlywedge} W_s=\delta_{rs} A-\epsilon_{rs}{}^t W_t~,~~~A\bar{\curlywedge} W_r=2 I_r-2J_r~,~~~
\cr
&&W_r\bar{\curlywedge} W_s=-2\epsilon_{rs}{}^t (I_t+J_t)~.
\label{comso5}
\eea
It turns out that $\mathfrak{f}=\Omega^2_{\hat\nabla}(M)=\mathfrak{so}(5)$.  This is not  unexpected. To see this, recall that for holonomy $Sp(2)\ltimes \bR^8$ backgrounds $\mathfrak{f}=\mathfrak{sp}(1)$. Note also that $\mathfrak{sp}(2)=\mathfrak{so}(5)$ and $\mathfrak{sp}(2)\oplus \mathfrak{sp}(1)$ is a maximal subgroup of $\mathfrak{so}(8)=\wedge^2\bR^8$. So if the invariant 2-forms under $\mathfrak{sp}(2)$ in $\wedge^2\bR^8$ span the Lie algebra $\mathfrak{sp}(1)$, then the invariant forms in $\wedge^2\bR^8$ under  (the holonomy Lie algebra) $\mathfrak{sp}(1)$ span $\mathfrak{so}(5)$.

To describe the Lie algebra structure of $\Omega^*_{\hat\nabla}(M)$, one has to determine the representation of $\mathfrak{f}$ on $\Omega^5_{\hat\nabla}(M)$.  For this observe that $\Omega^5_{\hat\nabla}(M)=\bR\langle L, M, A^2, P,  Q, N_{rs}, Y_r, Z_r\rangle$, where $L,M,N$ are defined in (\ref{LMN}), $A^2\equiv \curlywedge^2 A$,  $P\equiv {1\over3} \delta^{rs} I_r \curlywedge W_s$, $Q\equiv {1\over3} \delta^{rs} J_r \curlywedge W_s$ and $Y_r\equiv I_r \curlywedge A$ and $Z_r\equiv J_r \curlywedge A$.  To establish this note the identities
\bea
&&W_r\curlywedge W_s=(2\delta^{pq} N_{pq}+A^2)\delta_{rs}-N_{rs}-N_{sr}~,~~~I_r\curlywedge W_s=P\delta_{rs}-\epsilon_{rs}{}^t Y_t~,
\cr
&&J_r\curlywedge W_s=Q\delta_{rs}+\epsilon_{rs}{}^t Z_t~,~~~A\curlywedge W_r=-\epsilon_r{}^{st} N_{st}~.
\label{relso5}
\eea
It turns out that under the action of $\mathfrak{so}(5)$,  $\Omega^5_{\hat\nabla}(M)$ decomposes as
\bea
\Omega^5_{\hat\nabla}(M)=\Omega^5_{\bf 14}\oplus \Omega^5_{\bf 5}\oplus \Omega^5_{\bf 1}~,
\label{decomsp1}
\eea
where $\Omega^5_{\bf 14}$ is the irreducible representation of $\mathfrak{so}(5)$ constructed as the symmetric and traceless product of two vector representations,  $\Omega^5_{\bf 5}$ is the vector representation of $\mathfrak{so}(5)$ and the trivial representation $\Omega^5_{\bf 1}$ is spanned by the quadratic Casimir element of $\mathfrak{so}(5)$.  In particular, $\Omega^5_{\bf 5}=\bR\langle Y_r+Z_r, P+Q, L-M\rangle$. Apart from a direct calculation, this result can be established after decomposing $\wedge^4\bR^8$ in $(\mathfrak{sp}(1)\oplus \mathfrak{sp}(2))\subset \mathfrak{so}(8)$ representations. Then $\Omega^5_{\hat\nabla}(M)$ can be identified with the span of the elements in $\wedge^4\bR^8$, which are invariant under the action of $\mathfrak{sp}(1)$. Decomposing this subspace into $\mathfrak{sp}(2)$ representations yields (\ref{decomsp1}), see e.g. \cite{salamon} proposition 9.2 and references within.

It is worth pointing out that the form bilinears do not span   $\Omega^5_{\hat\nabla}(M)$. Indeed the 5-form bilinears are symmetric in the exchange of Killing spinors. As these backgrounds preserve five supersymmetries,   the 5-form bilinears span an at most 15-dimensional vector space, while the dimension of $\Omega^5_{\hat\nabla}(M)$ is 20.  Nevertheless, the additional elements of $\Omega^5_{\hat\nabla}(M)$ are $\hat\nabla$-covariantly constant and they should be included in the investigation as  they generate holonomy symmetries in sigma model action (\ref{act1}).

To determine the Lie algebra structure of $\Omega^*_{\hat\nabla}(M)$, it remains to compute $\Omega^5_{\hat\nabla}(M)\bar\curlywedge\Omega^5_{\hat\nabla}(M)$ using (\ref{LMNP}). This is a straightforward computation and the result follows upon application of (\ref{comso5}) and (\ref{relso5}). As the final formulae are not  illuminating, they will not be presented here.

\subsubsection{U(1)}

The Lie algebra of fundamental forms is $\mathfrak{f}=\mathfrak{u}(4)$.  This can be computed following the steps of a calculation similar to that performed for
$Sp(1)\ltimes\bR^8$ backgrounds in the previous section. As this is elaborate, we shall present instead an alternative group theoretic justification for the assertion. First, the Lie subalgebra $\mathfrak{u}(1)$ of the holonomy group, viewed as a 1-dimensional subspace of $\mathfrak{so}(8)=\wedge^2\bR^8$, is spanned by a complex structure $U$ on $\bR^8$. The elements of $\wedge^2\bR^8$ that are invariant under the action of $U$, and so of  $\mathfrak{u}(1)$, are the (1,1)-forms with respect to $U$.  It is known that the latter span the Lie algebra $\mathfrak{u}(4)$, which is the subalgebra of $\mathfrak{so}(8)$ that leaves invariant $U$.

As the 3-form bilinears are skew-symmetric in the exchange of the two Killing spinors and these backgrounds preserve six supersymmetries, they span an at most 15-dimensional subspace in $\Omega^3_{\hat\nabla}(M)$.  But as we have seen $\mathfrak{f}=\Omega^3_{\hat\nabla}(M)=\mathfrak{u}(4)$ has dimension 16.  The additional $\hat\nabla$-covariantly constant form is  represented by $U$ and it should be included in $\mathfrak{f}$ as it generates a holonomy symmetry for sigma model actions (\ref{act1}). The remaining symmetries are expected to be generated by form bilinears as $\wedge^2\bR^6=\mathfrak{so}(6)=\mathfrak{su}(4)$.

Similarly, the elements of $\Omega^5_{\hat\nabla}(M)$ are represented by the (2,2)-forms with respect to $U$ in $\wedge^4\bR^8$. Although, $\Omega^5_{\hat\nabla}(M)$ can be decomposed further into irreducible representations of $\mathfrak{u}(4)$ with this description of the action of $\mathfrak{f}$ on $\Omega^5_{\hat\nabla}(M)$ it suffices to specify the Lie algebra structure of $\Omega^*_{\hat\nabla}(M)$, as the action of $\mathfrak{u}(4)$ on (2,2)-forms is well known.

Again, $\Omega^5_{\hat\nabla}(M)$ contains more elements than those expected from counting 5-form bilinears. Indeed, the latter span an at most 21-dimensional space while $\Omega^5_{\hat\nabla}(M)$ has dimension 36. As it was mentioned in the previous case, all the elements of $\Omega^5_{\hat\nabla}(M)$ generate symmetries for the sigma model action (\ref{act1}) and so they should be included in the description.

All elements in $\Omega^5_{\hat\nabla}(M)$ can be written as  linear combinations of the $\curlywedge$-product of two elements in $\Omega^3_{\hat\nabla}(M)$.  Indeed as $(\bbe^{-\alpha\bar\beta}; \alpha,\beta=1,2,3,4)$   is a basis in $\Omega^3_{\hat\nabla}(M)$, $\bbe^{-\alpha_1\bar\beta_1}\curlywedge \bbe^{-\alpha_2\bar\beta_2}=\bbe^{-\alpha_1\bar\beta_1\alpha_2\bar\beta_2}$ span $\Omega^5_{\hat\nabla}(M)$.  As a result, one can apply the formula (\ref{LMNP}) to compute $\Omega^5_{\hat\nabla}(M)\bar\curlywedge\Omega^5_{\hat\nabla}(M)$.  One finds that
\bea
&&\bbe^{-\alpha_1\bar\beta_1\alpha_2\bar\beta_2}\bar\curlywedge \bbe^{-\alpha'_1\bar\beta'_1\alpha'_2\bar\beta'_2}=-\Big(\big(\delta^{\alpha_1\bar\beta_1'} \bbe^{-\bar\beta_1\alpha_2\bar\beta_2 \alpha_1'\alpha_2'\bar\beta'_2}-\delta^{\alpha_1\bar\beta_2'} \bbe^{-\bar\beta_1\alpha_2\bar\beta_2 \alpha_1'\alpha_2'\bar\beta'_1}\big)-(\alpha_2\leftrightarrow \alpha_1)\Big)
\cr
&&\qquad\qquad -\Big(\big(\delta^{\bar\beta_1\alpha_1'} \bbe^{-\alpha_1\alpha_2\bar\beta_2\bar\beta_1'\alpha'_2\bar\beta'_2}-\delta^{\bar\beta_1\alpha_2'} \bbe^{-\alpha_1\alpha_2\bar\beta_2\bar\beta_1'\alpha'_1\bar\beta'_2}\big)-(\bar\beta_2\leftrightarrow  \bar\beta_1)\Big)~.
\eea
This completes the description of the Lie algebra structure of $\Omega^*_{\hat\nabla}(M)$ and so that of the W-algebra commutators (\ref{walg}) of holonomy symmetries.

\subsubsection{$\{1\}$}

In this case it is clear that in a co-frame adapted to the $\bR^8$ holonomy,  $\Omega^*_{\hat\nabla}(M)$  is represented by the elements of the vector space $\wedge^*\bR^8$ including the even and odd degree forms. Keeping to the spirit of the discussion so far that $\Omega^*_{\hat\nabla}(M)$ has a Lie algebra structure, we shall restrict our analysis to the odd-degree $\hat\nabla$-covariantly constant forms.
In particular, the Lie algebra of fundamental forms $\mathfrak{f}=\Omega^3_{\hat\nabla}(M)$ is represented by elements of $\wedge^2\bR^8=\mathfrak{so}(8)$ and so $\mathfrak{f}=\mathfrak{so}(8)$. The elements of $\mathfrak{f}$ are spanned by 2-form bilinears.

Furthermore, $\mathfrak{f}=\mathfrak{so}(8)$ acts on $\Omega^5_{\hat\nabla}(M)$ with the standard representation on $\wedge^4\bR^8$. This is a reducible representation and decomposes into the sum $\wedge^{4+}\bR^8\oplus \wedge^{4-}\bR^8$ spanned by the self-dual and anti-self-dual 4-forms, respectively. Only the former represent 5-form bilinears.

Finally, as all elements of $\Omega^5_{\hat\nabla}(M)$ can be written as a linear combination of  $\curlywedge$-products of two elements in $\Omega^3_{\hat\nabla}(M)$, one can apply the formula (\ref{LMNP}) to compute $\Omega^5_{\hat\nabla}(M)\bar\curlywedge\Omega^5_{\hat\nabla}(M)$. Indeed as $(\bbe^{-ij}; i<j, i,j=1,\dots, 8 )$ is a basis in $\Omega^3_{\hat\nabla}(M)$ and $\bbe^{-ij}\curlywedge\bbe^{-kl}=\bbe^{-ijkl}$ span $\Omega^5_{\hat\nabla}(M)$, one finds  that
\bea
\bbe^{-ijkl}\bar\curlywedge\bbe^{-i'j'k'l'}= 4 (\delta^{[i|i'|} \bbe^{jkl]j'k'l'-}-\delta^{[i|l'|} \bbe^{jkl]i'j'k'-}+\delta^{[i|k'|} \bbe^{jkl]l'i'j'-}-\delta^{[i|j'|} \bbe^{jkl]k'l'i'-})~.
\eea
This,  together with the results in the beginning of section \ref{tpt}, complete the description of the Lie algebra structure of $\Omega^{\mathrm {odd}}_{\hat\nabla}(M)$ and so that of the commutators (\ref{walg}) of holonomy symmetries.
\vskip 0.5cm

Some of the results of  section 3 are summarised in the table below.
\begin{table}[h]
 \begin{center}
\begin{tabular}{|c||c|c|c|c|c|c|c|}\hline
$G$&$\mathrm{Spin }(7)$ & $SU(4)$& $Sp(2)$&$\times^2Sp(1)$&$Sp(1)$&$U(1)$&$\{1\}$
 \\ \hline
 $\mathfrak{f} $&\bR  & $\hat{\mathfrak{e}}(2) $ &$\mathfrak{sp}(1) $&$\oplus^2\mathfrak{sp}(1) $&$\mathfrak{so}(5) $&$\mathfrak{u}(4) $ &$\mathfrak{so}(8) $
\\ \hline
\end{tabular}
\end{center}
\vskip 0.2cm {\small Table 1: In the first row, the $G$ subalgebras of holonomy groups $G\ltimes\bR^8$ of the supersymmetric heterotic backgrounds are stated. In the second row, the associated Lie algebras of the fundamental forms are given. }

\end{table}

\section{Anomalies}

\subsection{Anomaly consistency conditions}

The consistency conditions for holonomy anomalies have been extensively investigated in \cite{lggpepb}. Here, we shall summarise without explanation the main formulae  and demonstrate that all holonomy anomalies of chiral sigma models on backgrounds with a non-compact holonomy group are consistent. To begin, the spacetime frame rotation (\ref{ftran2}) and gauge sector transformation (\ref{ftran1}) anomalies are given\footnote{For applications to string theory, replace $\hbar$ with $\alpha'$.} by
\bea
&&\Delta(\ell)={i\hbar\over 4\pi} \int d^2\sigma d\theta^+ Q^1_2(\omega, \ell)_{\mu\nu} D_+X^\mu \partial_=X^\nu~,~~
\cr
&&
\Delta(u)=-{i\hbar\over 4\pi} \int d^2\sigma d\theta^+ Q^1_2(\Omega, u)_{\mu\nu} D_+X^\mu \partial_=X^\nu~,
\label{lanuan}
\eea
respectively,
where the numerical coefficient in front of the expressions is determined after an explicit computation of relevant part  the effective action \cite{chpt2}.  The 2-form   $Q_2^1(\omega, \ell)$ is determined by the descent equations \cite{zumino} starting from the 4-form, $P_4(R)= \mathrm{tr} (R(\omega)\wedge R(\omega))$, which is proportional to the first Pontryagin form of the spacetime, as
\bea
 &&d P_4(R)=0~,~~\delta_\ell P_4(R)=0 \Longrightarrow   P_4(R)=dQ^0_3(\omega)~,~~
 \cr
 &&
 d\delta_\ell Q_3^0=0 \Longrightarrow \delta_\ell Q^0_3(\omega)=dQ^1_2(\ell, \omega)~,
 \eea
 and similarly for $Q^1_2(\Omega, u)$, where $R$ is the curvature of a frame connection $\omega$ of the spacetime.   The connection $\omega$ in the expression for the anomaly  is not uniquely specified as it can be altered with the addition of a suitable finite local counterterm in the effective action. This will be used later to prove the consistency of the anomalies.

 The anomaly consistency conditions \cite{lggpepb} of frame rotation and gauge sector anomalies (\ref{lanuan}) with the anomaly $\Delta(a_{{}_L})$  of a holonomy symmetry  generated by the $\hat\nabla$-covariantly constant form $L$ yield
 \bea
\Delta(a_{{}_L})={i\hbar\over 4\pi} \int d^2\sigma d\theta^+\, Q_3^0(\omega, \Omega)_{\mu\nu\rho} \delta_L X^\mu D_+X^\nu \partial_=X^\rho+ \Delta_{\mathrm {inv}}(a_{{}_L})~,
\label{Lanom}
\eea
where $ Q_3^0(\omega, \Omega)= Q_3^0(\omega)- Q_3^0(\Omega)$ and $\delta_\ell \Delta_{\mathrm {inv}}(a_{{}_L})=\delta_u \Delta_{\mathrm {inv}}(a_{{}_L})=0$. At one loop, we set $\Delta_{\mathrm {inv}}(a_{{}_L})=0$.  Furthermore, the consistency condition between two holonomy anomalies\footnote{It turns out that the above consistency condition is more general. If the anomaly of two transformations $\delta_1$ and $\delta_2$ is given as in (\ref{Lanom}), then their mutual consistency condition will be given as in (\ref{LMcon}) with $\delta_L=\delta_1$ and $\delta_M=\delta_2$.} generated by the $\hat\nabla$-covariantly constant forms $L$ and $M$ is
\bea
&&\delta_L \Delta(a_{{}_M})-\delta_M \Delta(a_{{}_L})= {i\hbar\over 4\pi} \int d^2\sigma d\theta^+ Q_3^0(\omega, \Omega)_{\mu\nu\rho} [\delta_L, \delta_M] X^\mu  D_+ X^\nu \partial_=X^\rho
\cr
&&\qquad\qquad +{i\hbar\over 4\pi} \int d^2\sigma d\theta^+ P_4(R, F)_{\mu\nu\rho\sigma} \delta_L X^\mu \delta_M X^\nu D_+ X^\rho
\partial_=X^\sigma
~,
\label{LMcon}
\eea
 where $P_4(R, F)=P_4(R)-P_4(F)$. It is clear that the consistency of two holonomy anomalies requires for the last term of (\ref{LMcon}) to vanish.

As supersymmetric backgrounds with non-compact holonomy group admit a null $\hat\nabla$-parallel vector field $K$,  the $\Delta(a_{{}_K})$ anomaly is consistent with any other holonomy anomaly $\Delta(a_{{}_M})$ provided that
 $i_K P_4(R, F)=0$.  Indeed $i_KP_4(F)=0$ as $i_K F=0$.  Therefore, it remains to choose the frame connection $\omega$ of the spacetime such that $i_KP_4(R)=0$.  Such a connection can always be found. For example, one can choose the frame connection  $\check \nabla$  whose torsion is $-H$.  Then, the result follows from the restriction of the holonomy of $\hat\nabla$ to be a subgroup of $G\ltimes \bR^8$ and the Bianchi identity $\hat R_{\mu\nu, \rho\sigma}=\check R_{\rho\sigma, \mu\nu}$ for $dH=0$.

Assuming that the connection $\omega$ is chosen such that $i_K P_4(R, F)=0$, it is straightforward to observe that the last term in (\ref{LMcon}) always vanishes for the holonomy transformations generated by  $\hat\nabla$-covariantly constant forms that are null along $K$.  This is a direct consequence of $i_K P_4(R, F)=0$ and the requirement that forms $L$ and $M$ are null along $K$. This can be verified by a short straightforward computation. As such, it can always be arranged  for  the anomalies of the holonomy symmetries to be consistent.

It should be stressed that the condition $i_K P_4(R, F)=0$ on $P_4(R, F)$ for the consistency of holonomy anomalies for backgrounds with a non-compact holonomy group is much  weaker than the analogous conditions for backgrounds with compact holonomy groups \cite{lggpepb} or those for manifolds with a G-structure \cite{dlossa}.  In particular, the application of the second condition in (\ref{invcon}) for the gauge sector and the analogous condition for the curvature of spacetime  are not required for the consistency of holonomy anomalies.

\subsection{Anomaly cancellation}

There are two ways to cancel the anomaly of holonomy symmetries. One is to add to the effective action  suitable finite local local counterterms to cancel the anomalies. This method has been successful in cancelling \cite{sen, phgpa}  the  second supersymmetry anomaly generated by a complex structure on the spacetime and requires a refinement of the Poincar\'e lemma and possibly the existence of special coordinates on the spacetime, such as, for example, the local triviality of Hodge cohomology and complex coordinates. This is not the case here. The forms $L$ that generate the holonomy symmetries, in general, are not associated with the existence of such structures in a straightforward manner. The second method is  to assume that the form $L$ that generates the holonomy symmetry is quantum mechanically corrected to $L^\hbar$, such that $L^\hbar$  is covariantly constant with respect to a connection $\hat \nabla^\hbar$ that has torsion
\bea
H^{\hbar}=H-{\hbar\over 4\pi} Q^0_3(\omega, \Omega)+{\cal O}(\hbar^2)~,
\label{corrh}
\eea
and $i_{L^\hbar} F^\hbar=0$. This is because the expression for the anomaly of the holonomy symmetry in (\ref{Lanom}) can be viewed as a correction to the covariant constancy condition on $L$ required for the transformation $\delta_L$ to be a symmetry of the action. Clearly, the new transformation $\delta_{L^\hbar}$ will be a symmetry of the effective action and the anomaly will cancel.  Such an anomaly cancellation mechanism is compatible with both the Green-Schwarz anomaly cancellation mechanism for gravitational anomalies \cite{mgjs}  and the heterotic supergravity effective action \cite{roo}.

For the cancellation of holonomy symmetry anomalies of sigma models on supersymmetric backgrounds with non-compact holonomy groups, the second cancellation mechanism is the most appropriate with the expectation that $L^\hbar$ remains null along $K$. The latter assertion is justified as it is not expected for the Killing spinors to be corrected up to two loops in sigma model perturbation theory, after choosing an appropriate co-frame on the spacetime. Alternatively, the classification of the geometry of heterotic supersymmetric backgrounds remains the same, up to two loop level, for backgrounds with either a closed or non-closed 3-form field strength $H$. In the former case, the condition $dH=0$ is imposed at the end. Note though that the anomaly of the symmetry generated by $K$ can be removed with the addition of  a finite local counterterm in the effective action. So $K$ need not to be corrected.

Furthermore, the above cancellation of the holonomy anomalies is consistent. The commutator of two holonomy symmetries for spacetimes with a non-compact holonomy group either vanishes or it closes to a type III transformation in the terminology of \cite{lggpepb}. In either case, the consistency condition is
\bea
P(\omega, \Omega)_{\mu\nu[\rho|\sigma|} L^\mu{}_L M^\nu{}_{M]}=0~.
\label{concona}
\eea
This is satisfied provided that the spacetime frame  connection $\omega$ is chosen such that $i_KP(\omega, \Omega)=0$. There  always exists such a connection in sigma model perturbation theory.

\section{Concluding remarks}

We have presented the W-algebra of holonomy symmetries of chiral sigma models propagating on supersymmetries heterotic backgrounds with non-compact holonomy group $G\ltimes \bR^8$. We have demonstrated that the W-algebra, which depends on $G\ltimes \bR^8$, is completely determined by a  Lie algebra   structure on the space of $\hat\nabla$-covariantly constant forms, $\Omega^*_{\hat\nabla}(M)$, which are null along a null vector field $K$.  The Lie on  bracket on $\Omega^*_{\hat\nabla}(M)$ is a generalisation of the  inner derivations on forms. Moreover, we identify the Lie algebra structure of $\Omega^*_{\hat\nabla}(M)$  for each holonomy group $G\ltimes \bR^8$ with $G$ given in (\ref{hol}).

In addition, we give the anomalies of the holonomy symmetries  up to and including 2-loops in sigma model perturbation theory. We demonstrate that they satisfy the Wess-Zumino consistency conditions, provided the frame connection is appropriately chosen on the spacetime. These anomalies are  cancelled, provided that the forms that generate the symmetries are corrected such that they remain null along $K$ and are covariantly constant with respect to a connection with torsion.  The torsion includes the difference of the Chern-Simons forms of the spacetime and  gauge sector connections of the sigma model.

It is remarkable that there is such an extensive Lie algebraic structure underpinning the W-algebra of holonomy symmetries associated with supersymmetric heterotic backgrounds with a non-compact holonomy group.  It is clear that this can be extended to all null forms along $K$, $\Omega^*_K(M)$, including those of even degree. The  bracket will again be given  by $\bar\curlywedge$. Of course $\Omega^*_K(M)$ is an infinite dimensional superalgebra where the odd degree forms span the Grassmannian even generators while the even degree forms span the Grassmannian odd generators of the superalgebra. This superalgebra structure can be restricted on
$\Omega^*_{\hat\nabla}(M)$. This is relevant in the investigation of holonomy symmetries of sigma models on heterotic supersymmetric backgrounds with holonomy $\bR^8$, as $\Omega^*_{\hat\nabla}(M)$  includes forms of both even and odd degree.
 Notice that $\Omega^*_{\hat\nabla}(M)$ contains a sub-superalgebra $\Omega^1_{\hat\nabla}(M)\oplus \Omega^2_{\hat\nabla}(M)\oplus \Omega^3_{\hat\nabla}(M)$, which can be seen as an extension of $\mathfrak{so}(8)=\Omega^3_{\hat\nabla}(M)$ with new Grassmannian odd
generators $\bR^8= \Omega^2_{\hat\nabla}(M)$ and central generator $\bR\langle K\rangle=\Omega^1_{\hat\nabla}(M)$.
These structures can be further extended to all forms on Euclidean and Lorentzian manifolds, where now the Lie (super)bracket is given by the standard inner derivation $\bar\wedge$.

The Lie algebraic structure that we have  uncovered   that underpins the W-algebra of holonomy symmetries for backgrounds with non-compact holonomy groups does not naturally extend to include that of backgrounds with compact holonomy groups investigated in \cite{lggpepb}, see also \cite{sven, gpph}.  This is because in the latter case the W-algebra can close to symmetries which are not generated by $\hat\nabla$-covariantly constant forms. For example, the W-algebra can close to worldsheet translations generated by the sigma model energy-momentum tensor. Nevertheless, it may be possible to extend the Lie (super)algebra structure on $\Omega^*_{\hat\nabla}(M)$ with additional generators that are not forms in such a way that the new algebra specifies the associated W-algebra of symmetries of sigma models.  It would be of interest to explore such a possibility in the future.

\section*{Acknowledgments}
JP is supported by the EPSRC grant EP/R513064/1.



\end{document}